\documentclass[aps,prx,reprint,showpacs,amsmath,amssymb,floatfix,superscriptaddress,noeprint]{revtex4-2}
\usepackage{bm}
\usepackage{color}
\usepackage[colorlinks,linkcolor=blue,urlcolor=blue,citecolor=blue,anchorcolor=blue]{hyperref}
\usepackage{graphicx}
\usepackage[capitalise]{cleveref}
\usepackage{physics}
\usepackage[detect-weight=true,separate-uncertainty=true]{siunitx}
\usepackage{multirow}

\begin{document}

\title{Dissipative stabilization of squeezing beyond \SI{3}{dB} in a microwave mode}

\author{R. Dassonneville}
\affiliation{Univ Lyon, ENS de Lyon, Univ Claude Bernard, CNRS, Laboratoire de Physique, F-69342 Lyon, France}
\author{R. Assouly}
\affiliation{Univ Lyon, ENS de Lyon, Univ Claude Bernard, CNRS, Laboratoire de Physique, F-69342 Lyon, France}
\author{T. Peronnin}
\affiliation{Univ Lyon, ENS de Lyon, Univ Claude Bernard, CNRS, Laboratoire de Physique, F-69342 Lyon, France}
\author{A. A. Clerk}
\affiliation{Pritzker School of Molecular Engineering, University of Chicago, Chicago, IL 60637, USA}
\author{A. Bienfait}
\affiliation{Univ Lyon, ENS de Lyon, Univ Claude Bernard, CNRS, Laboratoire de Physique, F-69342 Lyon, France}
\author{B. Huard}
\affiliation{Univ Lyon, ENS de Lyon, Univ Claude Bernard, CNRS, Laboratoire de Physique, F-69342 Lyon, France}

\date{\today}

\begin{abstract}
While a propagating state of light can be generated with arbitrary squeezing by pumping a parametric resonator, the intra-resonator state is limited to \SI{3}{dB} of squeezing. Here, we implement a reservoir engineering method to surpass this limit using superconducting circuits. Two-tone pumping of a three-wave-mixing element implements an effective coupling to a squeezed bath which stabilizes a squeezed state inside the resonator. Using an ancillary superconducting qubit as a probe allows us to perform a direct Wigner tomography of the intra-resonator state. The raw measurement provides a lower bound on the squeezing at about $\SI{6.7 \pm 0.2}{dB}$ below the zero-point level. Further, we show how to correct for resonator evolution during the Wigner tomography and obtain a squeezing as high as \SI{8.2 \pm 0.8}{dB}. Moreover, this level of squeezing is achieved with a purity of \SI{-0.4 \pm 0.4}{dB}. 
\end{abstract}

\maketitle
\section{Introduction}

One of the most striking predictions of quantum mechanics is that even in the ground state of an harmonic oscillator, any quadrature measurement is noisy. Zero point fluctuations can however be engineered and lowered for one quadrature of the field at the expense of the other. These squeezed states have become a central resource for quantum information processing. They can be used to boost the sensitivity of many measurements including gravitational wave detection~\cite{Kimble2001,Abadie2011,Aasi2013,Korobko2017}, perform quantum secure communication~\cite{DrummondBook,Braunstein2005} and used for measurement-based continuous-variable quantum computing \cite{Braunstein2005,Zhong2020}. Squeezing is usually generated by parametrically pumping a resonator. This process generates squeezing of both the intra-resonator and outgoing fields. While any amount of squeezing can theoretically be obtained for the outgoing field, the steady-state intra-resonator squeezing is limited to \SI{3}{dB} below the zero point fluctuations. 

Intra-resonator squeezing beyond \SI{3}{dB} can in principle be attained by injecting squeezed light into the resonator input using an external source of squeezed radiation~\cite{Kono2018,Eddins2018,Malnou2019}.  In practice however, the achievable squeezing in such schemes is limited by losses associated with transporting and injecting the extremely fragile squeezed state into the resonator.  
A more attractive approach is to use reservoir-engineering techniques~\cite{Poyatos1996}, where tailored driving results in the cavity being coupled to effective squeezed dissipation~\cite{CiracParkins1993,Kronwald2013,Didier2014}. These methods can also surpass the \SI{3}{dB} limit, and do not involve transporting an externally-prepared squeezed state.  Reservoir-engineering intracavity squeezing beyond  
\SI{3}{dB} has recently been achieved for mechanical modes, both in optomechanical systems~\cite{Wollman2015,Pirkkalainen2015,Lecocq2015,Lei2016} as well as in a trapped ion platform~\cite{Kienzler2015}.  

In this work, we experimentally demonstrate that reservoir-engineering squeezing beyond \SI{3}{dB} can also be achieved for purely electromagnetic intracavity modes, namely a microwave-frequency mode in a superconducting quantum circuit. Using the well developed circuit-QED toolbox, we also perform a direct tomography of the intra-resonator squeezed state instead of inferring the resonator state from the measured output mode.  
This is achieved through the use of 
an ancillary superconducting qubit, which enables \textit{in-situ} Wigner tomography of the squeezed intracavity microwave mode. The intracavity squeezing factor reaches at least \SI{-6.7 \pm 0.2}{dB}, going well beyond the \SI{3}{dB} limit.  We also probe the non-classicality of the squeezed state by investigating its photon number statistics~\cite{Kono2017}, and use our tomographic method to carefully study the full dynamics of the dissipative generation of squeezing.  
This work thus presents an interesting platform to stabilize, manipulate and characterize Gaussian states in-situ. Our stabilization technique could also be extended beyond simple squeezed states to other continuous variable states such as cat or grid states~\cite{Puri2017,Grimm2019,Campagne-Ibarcq2020,Fluhmann2019,Hastrup2020,Neeve2020} by taking advantage of the large non-linearities that can be engineered in circuit-QED. 

\section{System and model}
\begin{figure}
\centering
\includegraphics[width=8.6cm]{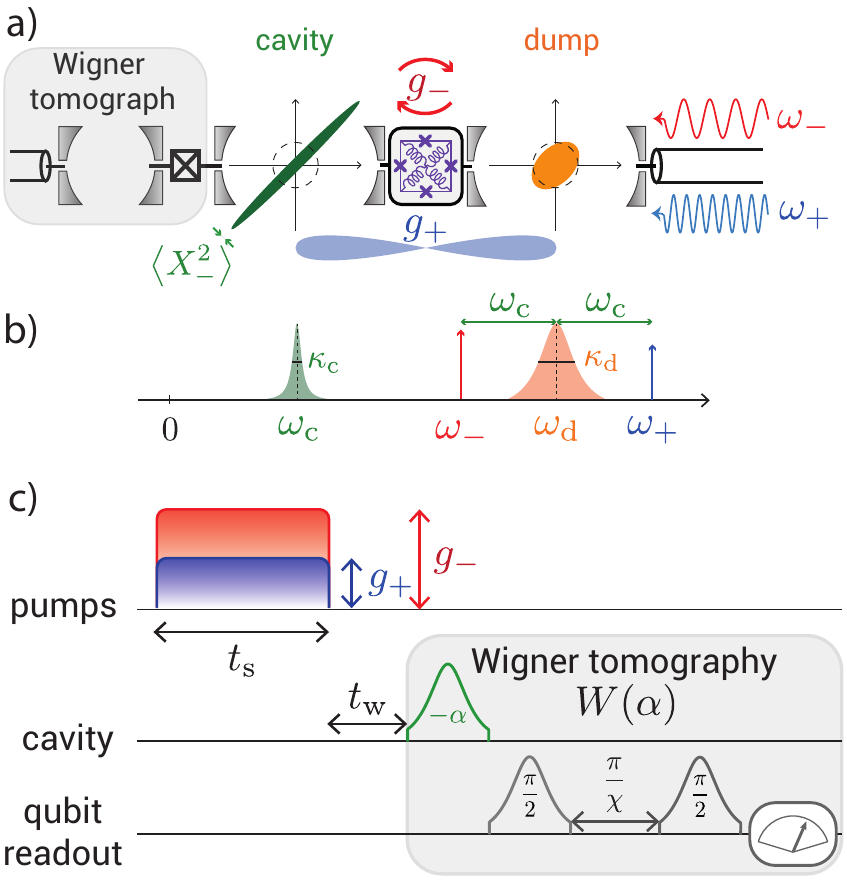}
\caption{a) Principle of the experiment. A cavity mode (green) at frequency $\omega_\mathrm{c}$ is coupled to a 
dump mode at frequency $\omega_{\mathrm{d}}$ (orange) via a Josephson Ring Modulator (JRM, in purple). The dump mode is strongly coupled to a cold transmission line through which the JRM is pumped at both frequencies $\omega_+ = \omega_\mathrm{c}+\omega_\mathrm{d}$ (two-mode squeezing) and $\omega_- = \omega_\mathrm{d}-\omega_\mathrm{c}$ (photon conversion). A squeezed vacuum state is stabilized into the cavity as a result. An ancillary qubit with an ancillary readout resonator is used as a Wigner tomograph. The contours of the Wigner functions of each mode are shown as colored regions in the quadrature phase space, while a dashed circle represents the vacuum state. b) Frequencies of the involved modes and drives.  c) Pulse sequence. The sum pump at $\omega_+$ with amplitude $g_+$ and the difference pump at $\omega_-$ with amplitude $g_-$ are applied for a time $t_\mathrm{s}$. After a waiting time $t_\mathrm{w}$, the Wigner function of the cavity $W(\alpha)$ is measured using a cavity displacement by $-\alpha$ followed by a parity measurement~\cite{Lutterbach1997,Bertet2002,Vlastakis2013}. } 
\label{fig:principle}
\end{figure}

Our device consists in a Josephson Ring Modulator (JRM)~\cite{Bergeal2010} coupling one mode (the \emph{cavity}) which we would like to stabilize in a squeezed state, and a second auxiliary mode strongly coupled to a transmission line (the \emph{dump}). The cavity and dump have resonant frequencies $\omega_{\rm c}/2\pi = \SI{3.74155}{GHz}$ and $\omega_{\rm d}/2\pi = \SI{11.382}{GHz}$ and decay rates $\kappa_{\rm c}/2\pi = \SI{40}{kHz}$ and $\kappa_{\rm d}/2\pi = \SI{8}{MHz}$. Our setup also has an ancillary transmon qubit coupled to the cavity; its only role is to perform intra-resonator Wigner tomography (\cref{fig:principle}.a). 

When applying a pump at frequency $\omega_\mathrm{-} = \omega_{\rm d} - \omega_{\rm c}$, and within the rotating-wave approximation (RWA) and stiff pump condition, the JRM leads to a beam-splitter interaction Hamiltonian $\hat{H}_\mathrm{-}/\hbar = g_{\rm -} \hat{d}^{\dag} \hat{c} + g_{\rm -}^* \hat{d} \hat{c}^{\dag}$, where the pump amplitude controls the coupling strength $g_-$ between the cavity and dump modes described by bosonic operators $\hat{c}$ and $\hat{d}$. It mediates coherent exchange of photons between the cavity and the dump and thus lossless frequency conversion~\cite{Bergeal2010b,Abdo2013a}. In contrast, a pump applied at frequency $\omega_+ = \omega_{\rm d} + \omega_{\rm c}$ mediates a parametric down conversion process involving cavity and dump,  $\hat{H}_\mathrm{+}/\hbar =  g_{\rm +} \hat{d}^{\dag} \hat{c}^{\dag} + g_{\rm +}^* \hat{d} \hat{c}$.  The pump amplitude controls the coupling strength $g_+$. On its own, this kind of pumping leads to phase-preserving amplification~\cite{Bergeal2010,Roch2012} and generation of two-mode squeezed states~\cite{Flurin2012}. Note that in order to avoid parasitic nonlinear effects, we operate the JRM at a flux point which maximizes these three-wave mixing terms while cancelling the four-wave mixing terms~\cite{Peronnin2019,dassonneville2020b}.
\begin{figure*}
\centering
\includegraphics{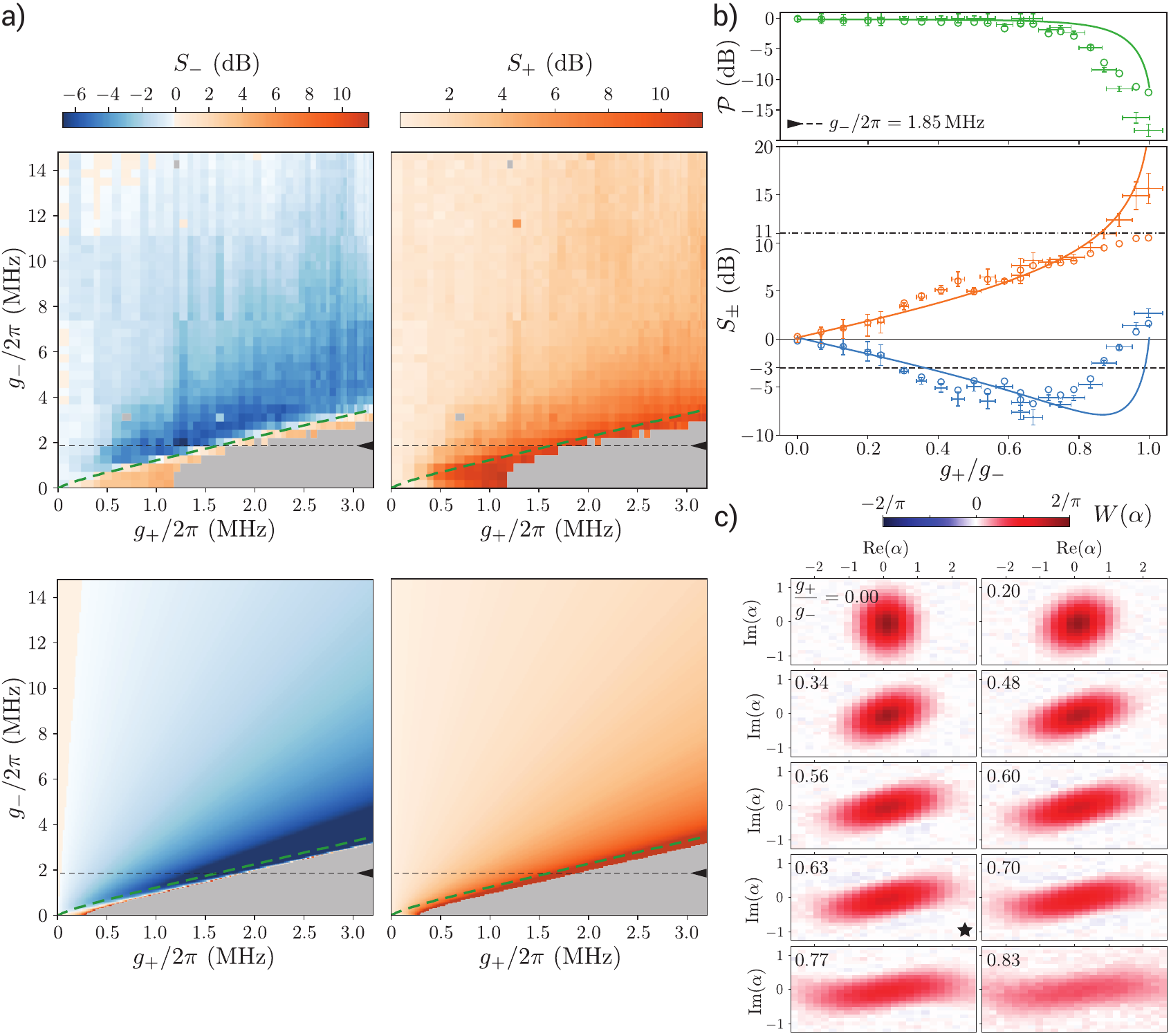}
\caption{Characterization of the stabilized squeezed state. a) Top panels: measured steady-state squeezing $S_- = \expval{X_-^2}/X_0^2$ (left) and anti-squeezing $S_+ = \expval{X_+^2}/X_0^2$ (right) factors. Bottom panels: theoretical prediction for $S_\pm$ using \cref{eq:analytical_variance}. 
Green dashed lines correspond to the value $g_-^{\mathrm{opt}}$ as a function of $g_+$ that minimizes the squeezing $S_-$ according to \cref{eq:analytical_variance} (i.e.~calculated by neglecting Kerr nonlinearities).
b) Purity $\mathcal{P}$ (top, green), squeezing $S_-$ (bottom, blue) and anti-squeezing $S_+$ (bottom, orange) factors as a function of $g_+/g_-$ for a fixed value $g_-/2\pi = \SI{1.85}{MHz}$ (cut along the arrow in Fig.~a). Circles are the normalized eigenvalues of the covariance matrix of the measured Wigner functions at each pump amplitudes as shown in (c) and reach a squeezing factor as low as $S_- = \SI{-6.7 \pm 0.2}{dB}$.
Points with error bars are the values obtained when correcting for cavity evolution during Wigner tomography (see~\cref{append:wigner_evolution}), which reveals a stabilized squeezing reaching as low as $S_- = \SI{-8.2 \pm 0.8}{dB}$. Solid lines come from the model \cref{eq:analytical_variance}. 
c) Selected measured Wigner functions along the same axis $g_-/2\pi = \SI{1.85}{MHz}$, for various $g_+/g_-$ ratios as indicated in the labels. The star indicates the Wigner function at optimum squeezing.}
\label{fig:steadystate}
\end{figure*}

Simultaneously pumping at these two frequencies enables various interesting phenomena such as effective ultrastrong coupling~\cite{Fedortchenko2017,Markovic2018} or directional amplification~\cite{Metelmann2017,Metelmann2015,Chien2020}. Here, using a long-lived cavity mode, we show that this double pumping scheme can stabilize a squeezed state \cite{CiracParkins1993,Kronwald2013}. Indeed, in the rotating frame, and setting the phase references such that $g_\pm$ are positive, the total Hamiltonian reads \begin{align}
    \hat{H}/\hbar =&  \hat{d}(g_{\rm +} \hat{c} + g_{\rm -} \hat{c}^{\dag} ) + h.c.  
\end{align}

In the case where $g_\mathrm{+}<g_\mathrm{-}$, this Hamiltonian can be reinterpreted as a beam splitter interaction between the dump mode and a Bogoliubov mode $\hat{\beta} = \cosh(r) \hat{c} + \sinh(r) \hat{c}^{\dag}$ with $r = \tanh^{-1}(g_{\rm +}/g_{\rm -})$. It reads
\begin{align}
    \hat{H}/\hbar = \mathcal{G} \hat{d}\hat{\beta}^{\dag}  + h.c., \label{eq_beam_splitter}
\end{align}
where the coupling strength is $\mathcal{G} = \sqrt{g_{\rm -}^2 - g_{\rm +}^2}$. In the ideal case where the coupling rate $\kappa_d$ of the dump mode to a reservoir at zero temperature is much larger than any other rates, and where the cavity lifetime $\kappa_c^{-1}$ is unlimited, the Hamiltonian leads to the relaxation of the Bogoliubov mode into its ground state. In that state, the cavity mode is a vacuum squeezed state with squeezing parameter $r = \tanh^{-1}(g_{\rm +}/g_{\rm -})$. 

The signature of this squeezing is best seen in the quadrature phase space of the cavity mode. We denote $X_-$ and $X_+$ the quadratures of the cavity mode that have the smallest and largest variances in a given state. In the vacuum state of the cavity ($r=0$), the variance of the quadratures corresponds to the zero point fluctuations $\expval{X_{\pm}^2 }_{|0\rangle}=X_0^2$. The squeezing factor one can generate in the ground state of the Bogoliubov mode is simply a scaling of the variances by the factor $S_{\pm}=\expval{X_{\pm}^2 }/X_0^2 = e^{\pm 2r}$. In the general case, where the Bogoliubov mode is not cooled down to its ground state, these factors become~\cite{Kronwald2013}

\begin{align}
    S_{\pm} = e^{\pm 2r} \expval{(\beta \mp \beta^\dagger)^2}.
    \label{eq:ss_sqz_beta}
\end{align}

We thus see that in principle,  the \SI{3}{dB} squeezing limit can be surpassed arbitrarily by 
having $g_\mathrm{+}$ approach $g_\mathrm{-}$ from below (as this causes the squeezing parameter $r$ to diverge).  
However, in this limit the effective coupling rate $\mathcal{G}$ of the Bogoliubov mode to the dump goes down to zero. As a result, the competition between this engineered decay channel and the intrinsic cavity loss (rate $\kappa_c$) prevents the Bogoliubov mode from reaching its ground state.  This both degrades the effective squeezing of the steady state, as well as its purity. Thus, for any value of $g_-$ there exists an optimum value of $g_+$ that minimizes the variance $\expval{X^2_\mathrm{-}}$. This minimum increases with the value of $g_-$ and is finally expected to saturate to a level set by the damping rates $S_{-} \geq \kappa_\mathrm{c}/(\kappa_\mathrm{c}+\kappa_\mathrm{d})$, which reflects the fact that the damping rate of the dump $\kappa_\mathrm{d}$ sets an upper limit to the coupling of the bosonic mode to the effective squeezed reservoir.

We thus see that a prerequisite for achieving squeezing well beyond \SI{3}{dB} is to engineer a large ratio $\kappa_\mathrm{d}/\kappa_\mathrm{c}$. For our sample parameters, we have $\kappa_\mathrm{d} \simeq 200 \kappa_\mathrm{c}$, leading to a lower bound of $S_{-}\geq \SI{-23}{dB}$~\cite{Kronwald2013}. Further, taking into account the thermal equilibrium occupancies $n^{\rm th}_{\rm c}$ and  $n^{\rm th}_{\rm d}$ of the cavity and dump modes, and in the limit of the experiment where $\kappa_c, \mathcal{G} \ll \kappa_d$, \cref{eq:ss_sqz_beta} leads to (see \cref{append:theory_sqz} or~\cite{Lei2016} for formula without approximation)  
\begin{align}
    S_{\pm}\simeq \frac{ \kappa_{\rm c} (2n^{\rm th}_{\rm c} +1) + \Gamma_{\pm}( 2 n_{\rm d}^{\rm th} +1) }{\kappa_{\rm eff}}
\label{eq:analytical_variance}
\end{align}
where we introduce $\kappa_{\rm eff} = \kappa_{\rm c} + 4\mathcal{G}^2/\kappa_{\rm d}$ and $\Gamma_{\pm} = 4 (g_{\rm -} \pm g_{\rm +})^2/\kappa_{\rm d}$.
 \cref{eq:analytical_variance} also makes it clear that the intrinsic loss rate $\kappa_c$ and non-zero environmental temperatures also lower the purity of the steady state $\mathcal{P}= \Tr(\rho^2)$ below $1$, where $\rho$ is the steady state cavity density matrix. This follows from the fact that $\mathcal{P}= 1/\sqrt{S_- S_+}$ for a Gaussian state.  

To measure the squeezing and anti-squeezing factors $S_\pm$, we perform a full \textit{in-situ} Wigner tomography~\cite{Lutterbach1997,Bertet2002,Vlastakis2013} using an ancillary transmon qubit at frequency $\omega_q/2\pi= \SI{4.32731}{\giga\hertz}$ (\cref{fig:principle}.a). It couples dispersively to the cavity with a dispersive shift $\chi/2\pi = \SI{-3.28}{\mega\hertz}$. A third resonator, at frequency $\omega_r/2\pi= \SI{6.293}{\giga\hertz}$, is used to perform single-shot readout of the qubit state with a fidelity of \SI{96}{\%} in a \SI{380}{ns} integration time.
From the Wigner function, we compute the covariance matrix of the cavity mode quadratures and diagonalize it to extract the minimum and maximum cavity quadrature variances $\langle X_\pm^2\rangle$. Due to its coupling to the qubit, the cavity acquires an induced parasitic self-Kerr nonlinearity $-K \hat{c}^{\dag^{2}} \hat{c}^{^{2}}$ and a qubit-state-dependent self-Kerr $-K_e \hat{c}^{\dag^2}\hat{c}^{^2} \ketbra{e}$ with $K/2\pi = \SI{20}{kHz}$ and $K_e/2\pi = \SI{70}{kHz}$ (measured in a previous run of the experiment). 
These non-linearities distort the squeezed state and thus reduce the effective squeezing factor, similarly to what occurs for Josephson parametric amplifiers (JPA)~\cite{Boutin2017}. While no analytical solution taking into account the Kerr effects exists, \cref{eq:ss_sqz_beta} and \cref{eq:analytical_variance} still provide a good description when $\mathcal{G} \gg K$. In the future, these non-linearities could be harnessed as a resource to stabilize more complex non-Gaussian states~\cite{Goto2016,Puri2017,Mamaev2018,Grimm2019}. 

\section{Steady-state squeezing}

The key advantage of reservoir engineering is that the desired target state is prepared in the steady state, independent of the initial cavity state: one can simply turn on the pumps and wait. We thus turn on $g_+$ and $g_-$ for a duration $t_\mathrm{s} = \SI{4}{\mu s}$ (cf \cref{fig:principle}.c) that is long enough to establish a steady state, and immediately afterwards measure the Wigner function $W(\alpha)$. To perform the measurement at each amplitude $\alpha$, we start by applying a calibrated displacement $D(-\alpha)$ to the cavity state using a cavity drive at $\omega_c$ with a pulse shape chosen to be a \SI{13}{ns} wide hyperbolic secant and whose complex amplitude is proportional to  $-\alpha$. We then measure the cavity parity operator by reading out the qubit state after performing two $\pi/2$ unconditional pulses on the qubit separated by a waiting time $\pi/\chi = \SI{152}{ns}$. We perform phase-cycling, running each sequence twice with an opposite phase for the second $\pi/2$ pulse, so as to remove most of the parasitic contribution of higher order Kerr effects~\cite{Kirchmair2013}. The Wigner function is probed on a discretized phase space using a rectangular grid of \SI{25x25}{pixels} approximately aligned  to the squeezing axis. Due to the finite window size (cf \cref{append:finite_window}), we could only resolve anti-squeezing up to $\SI{11}{dB}$ (dotted dash line in \cref{fig:steadystate}.b). Each Wigner tomogram is averaged over \SI{5000}{} realizations. 
To increase the repetition rate and limit the low-frequency drifts, the cavity is first emptied by applying a difference pump $g_-$, cooling it down to a thermal vacuum state with residual population $n_\mathrm{c}^\mathrm{th} = n_\mathrm{d}^\mathrm{th} =0.017 \pm 0.003 $ (see \cref{append:calib_nth}). The qubit is also reset to its ground state using measurement-based feedback.
Furthermore, to minimize the low-frequency noise as much as possible, we interleave pump-on-measurements with pump-off-measurement. We thus obtain experimental squeezing factors $S_\pm = \left(\expval{X^2_\pm}/\expval{X^2}_\mathrm{off}\right) \cdot \left( \expval{X^2}_\mathrm{off}/X_0^2\right)$ by first normalizing the measured variances $\expval{X_{\pm}^2}$ with the measured pumps-off variances $\expval{X^2}_\mathrm{off}$ and then correcting for the thermal occupancy $\expval{X^2}_\mathrm{off}/X_0^2 = \SI{0.15 \pm 0.03}{dB}$. 
The pumping strengths $g_+$ and $g_-$ are calibrated using independent measurements (\cref{append:pumps_calib}). We estimate a statistical uncertainty of \SI{\pm 0.2}{dB} on the variances extracted from the measured Wigner functions.

The obtained steady-state squeezing and anti-squeezing factors are displayed in \cref{fig:steadystate}.a) as a function of $g_+$ and $g_-$. We observe a maximum squeezing of \SI{-6.7 \pm 0.2}{dB} well below the \SI{-3}{dB} limit, which we believe to be the highest squeezing factor observed in an intracavity microwave mode. Correspondingly, we extract an anti-squeezing of \SI{7.7 \pm 0.2}{dB} and thus a state purity of \SI{-0.5 \pm 0.2}{dB}.  
For each value of the rate $g_+$, the largest squeezing we observe occurs for $g_- $ close to $g_+$. This trend is expected from the analytical Kerr-free model \cref{eq:analytical_variance} of the system, which predicts the working point of largest squeezing as a function of $g_+$ (green dashed line in \cref{fig:steadystate}.a)~\cite{Kronwald2013}. However, contrary to the expected monotonic increase of the optimal squeezing factor $S_-$ with $g_+$ (Kerr-free prediction in bottom panels of \cref{fig:steadystate}.a), we find a global maximum squeezing at a finite value of $(g_-,g_+)$. Note that for $g_+ > g_-$ the system becomes unstable: the qubit gets ionized~\cite{Lescanne2019b}, preventing us from measuring the Wigner functions (grey shade area). 

In \cref{fig:steadystate}.b-c), we show the squeezing and anti-squeezing as a function of $g_+/g_-$ as well as some measured Wigner tomograms, for $g_-/2\pi = \SI{1.85}{MHz}$. For $g_+<0.7 g_-$, the measured variances are well captured by \cref{eq:analytical_variance} with exponentially increasing squeezing factors. For $g_+>0.7g_-$, the measured variances start deviating from the theory (solid lines in \cref{fig:steadystate}.b). As can be seen in \cref{fig:steadystate}.c), the squeezed states are not Gaussian anymore in this parameter region. The Wigner functions develop an S-shape, a typical signature of the cavity self-Kerr. We attribute this effect to higher order terms we have so far neglected: the self-Kerr rate induced by the qubit on the cavity, as well as a residual four-wave mixing term in the JRM Hamiltonian (see \cref{append:num_model}). 

While the raw measurement of the Wigner function provides a good estimate of the steady-state squeezing parameter (circles in \cref{fig:steadystate}.b), the finite measurement time needed to perform tomography leads to a systematic error.  During this finite measurement time, the pump tones $g_{\pm}$ are off, implying that the cavity is no longer coupled to an effective squeezed reservoir.  The squeezed state thus degrades due to the intrinsic cavity loss.  A further error is caused by evolution under the cavity-self Kerr nonlinearity during this time. Both these effects cause our Wigner function methods to {\it underestimate} the true value of the steady-state squeezing.    

It is possible to correct for this measurement error and retro-predict via numerical simulation the squeezing factors $S_\pm$ associated with the state prepared at the end of the stabilization period~\cite{PhysRevA.93.012109}. To that end, we consider a series of input model Gaussian states for which we numerically implement our experimental Wigner tomography measurement. At the end of these simulations, we obtain a mapping from Gaussian states to measured squeezing and anti-squeezing factors that we are able to invert in order to retro-predict the stabilized state (\cref{append:wigner_evolution}). Using this correction improves the best squeezing estimate to \SI{-8.2 \pm 0.8}{dB} (dots with error bars in \cref{fig:steadystate}.b) with purity \SI{-0.4 \pm 0.4}{dB}.

It is interesting to compare our stabilization technique to other intra-resonator microwave squeezing generation schemes. One possibility consists in driving a cavity with a squeezed input state that is externally generated by a Josephson parametric amplifier (JPA)~\cite{Malnou2018,Malnou2019}. High squeezing factors~\cite{Mallet2011,Boutin2017,Malnou2018,Pogorzalek2019} ($\simeq \SI{-10}{dB}$) can be achieved in the amplifier output field.  However, transferring this state into a cavity is challenging as it is extremely sensitive to microwave losses, resulting in degraded squeezing and purity.
For comparison, we consider a resonator driven by a pure squeezing source (in practice a JPA). To achieve the same intracavity squeezing and purity as our setup
($S_-=\SI{-8.2}{dB}$ and $\mathcal{P}=\SI{-0.4}{dB}$ respectively), the source would need to generate an output squeezing better than $\sim$\SI{-9.1}{dB} and the losses between the source and the resonator would need to be kept below \SI{0.15}{dB}. 
This level of loss is smaller than the typical insertion loss of common microwave components. It is hard to achieve, even if all elements are fabricated in a single-chip architecture. We can also compare against another 
approach for generating (but not stabilizing) a squeezed state, based on the use of
arbitrary state preparation techniques (e.g.~the SNAP gate protocol~\cite{Heeres2015}). In Ref.~\cite{Wang2017}, the authors used such an approach to obtain a squeezing factor of \SI{-5.71}{dB} for a purity of \SI{-0.86}{dB} with a non-deterministic success rate of \SI{15}{\%}.

\section{Nonclassical photon distribution}
One of the hallmarks of vacuum squeezed states is that they are quantum superpositions involving only even-number photon Fock states.  Such ideal states have the form $| \psi \rangle = \sum_{k=0}^{\infty} \tanh(r)^k \frac{\sqrt{(2k)!}}{2^k k!} \ket{2k}$.  Our device allows us to directly verify this unique, non-classical aspect of the squeezed states we stabilize in our cavity~\cite{Kono2017}. This is because our coupling to the ancilla qubit is strong enough to place us in photon-number-resolved regime where distinct cavity photon numbers can be resolved by measuring the effective qubit frequency, i.e.~$\chi \gg \Gamma_2$ with $\Gamma_2 = (\SI{11}{\mu s})^{-1}$ the qubit coherence rate.

\begin{figure}
\centering
\includegraphics[width=8.6cm]{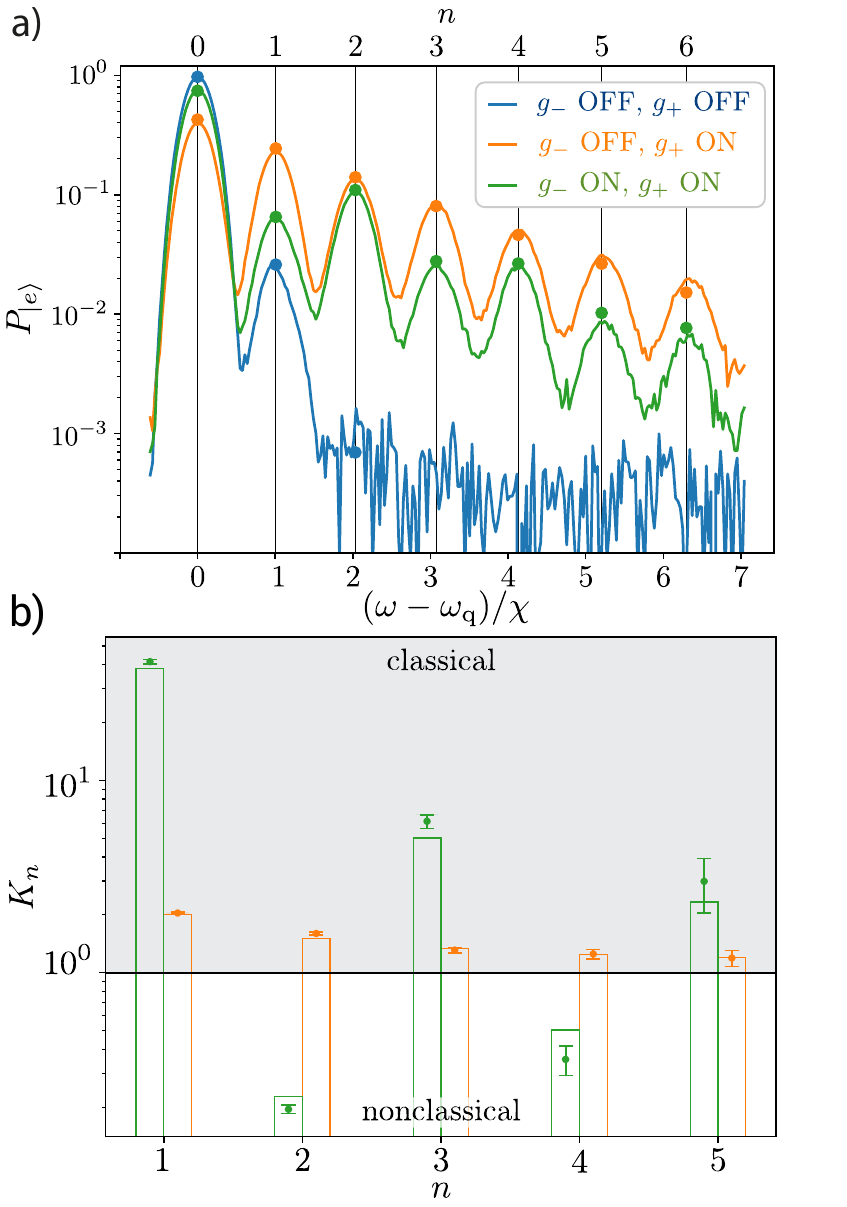}
\caption{a) Number photon distribution measurement using qubit spectroscopy. Solid lines: measured probability $P_{\ket{e}}$ that the qubit gets excited by a \SI{200}{ns} wide hyperbolic secant $\pi$-pulse of frequency $\omega$ after a cavity state is stabilized. Blue: near vacuum state when no pumps are applied. Orange: thermal state when only a pump $g_+$ is applied. Green: squeezed vacuum state when both pumps $g_+$ and $g_-$ are applied. Vertical lines indicate the qubit resonance frequency conditioned on the cavity having $n$ photons. Filled circles: numerical simulations. b) Dots with error bars: Klyshko number $K_n$ (see main text) calculated from the qubit spectroscopy. Orange bars: expected Klyshko number for a thermal state at any temperature. Green bars: Klyshko numbers predicted with the model described in the text.}
\label{fig:photon_distrib}
\end{figure}

After preparing the squeezed state, we perform spectroscopy of the qubit using a narrow-bandwidth $\pi$-pulse at a varying probe frequency $\omega$ followed by qubit readout (green curve in \cref{fig:photon_distrib}.a). The observed peak heights at each frequency $\omega-\omega_q\approx n\chi$ allow us to determine the cavity photon-number distribution $\mathbb{P}(n)$~\cite{Schuster2007}. We correct this dataset for the qubit residual thermal population (1 \%), the finite fidelity of the $\pi$-pulse and readout errors. Interestingly, the peaks are not evenly spaced in frequency due to the higher nonlinear term $-K_e \hat{c}^{\dag^2}\hat{c}^{^2} \ketbra{e}$ with $K_e/2\pi = \SI{70}{kHz}$. The photon number distribution $\mathbb{P}(n)$ are then obtained from the qubit excitation probability at $\omega_q - n (\chi + 2K_e n) $ (vertical lines). For comparison, we also measure the photon number distribution $\mathbb{P}(n)$ for two other cavity states: a thermal equilibrium state when no pumps are applied (blue in \cref{fig:photon_distrib}) and a thermal state that we create by only applying a sum pump $g_{\rm +}/2\pi = \SI{0.43}{MHz}$ (orange curve). Note that this thermal state is obtained by tracing out the dump mode for the vacuum two mode squeezed state that is stabilized between cavity and dump~\cite{flurin2015superconducting}.  

For the squeezed state (green curve with $g_{\rm -}/2\pi = \SI{2.2}{MHz}$, $g_{\rm +}/2\pi = \SI{1.42}{MHz}$), we observe a non-monotonic behavior: the weight of even photon numbers is enhanced, whereas that of odd photon numbers is suppressed (note the log scale here). The non-zero but small population of odd Fock states indicates a deviation from an ideal squeezed vacuum state. The measured data closely fits to our numerical simulation (dots in \cref{fig:photon_distrib}.a).
The measurement done with the pumps off gives the thermal population of the cavity $n_\mathrm{c}^\mathrm{th} = 0.017$ (blue dots) and also indicates the measurement noise floor. For the thermal state, we observe a Bose-Einstein distribution with a population $n_\mathrm{c}^\mathrm{th} = 1.5$ (orange dots).

Even though their Wigner function is always positive, squeezed states are typically regarded as non-classical states as they cannot be represented as statistical mixtures of coherent states. Formally, this means that they do not have well-behaved Glauber-Sudarshan $P$ representations~\cite{Klyshko1996}. Equivalently, it also manifests itself in the behaviour of so-called Klyshko numbers $K_n = (n+1) \mathbb{P}(n-1)\mathbb{P}(n+1) / n \mathbb{P}(n)^2$.
A state is non-classical if for one or more integers $n$, $K_n < 1$ 
(as this implies that the $P$ function cannot be well behaved). For example, a perfect squeezed vacuum state, as it only includes even photon numbers, exhibits infinite odd Klyshko numbers and zero even Klyshko numbers. Ref.~\cite{Kono2017} computed the Klyshko numbers for a squeezed state generated by an external JPA and observed a Klyshko number smaller than $1$, even though they only observed monotonic behavior in the photon distribution $\mathbb{P}(n)$.

For a thermal state, the Klyshko number are given by $K_n^\mathrm{th} = (n+1) / n$ independently of temperature (orange bars in \cref{fig:photon_distrib}.b). We observe this universal relation with the prepared thermal state (orange points with errorbar). Interestingly, it is a striking demonstration of the fact that a two mode squeezed state generates a thermal distribution when tracing out one of the modes. It is expected from the maximally entangled state at a given average energy. We do not show the Klyshko numbers when the pumps are off because $\mathbb{P}(n)$ is below the noise floor.

For the squeezed state, we observe ample oscillations in the Klyshko numbers (notice the log scale again). We measure $K_2=0.23$ and $K_4=0.5$ that are well below one (green points with errorbars). Similarly to the cavity population $\mathbb{P}(n)$, our numerical model (green bars) reproduces the observed Klyshko numbers.

\section{Stabilization dynamics and decay of squeezing}
\label{sec:dynamics}
\begin{figure}
\centering
\includegraphics[width=8.6cm]{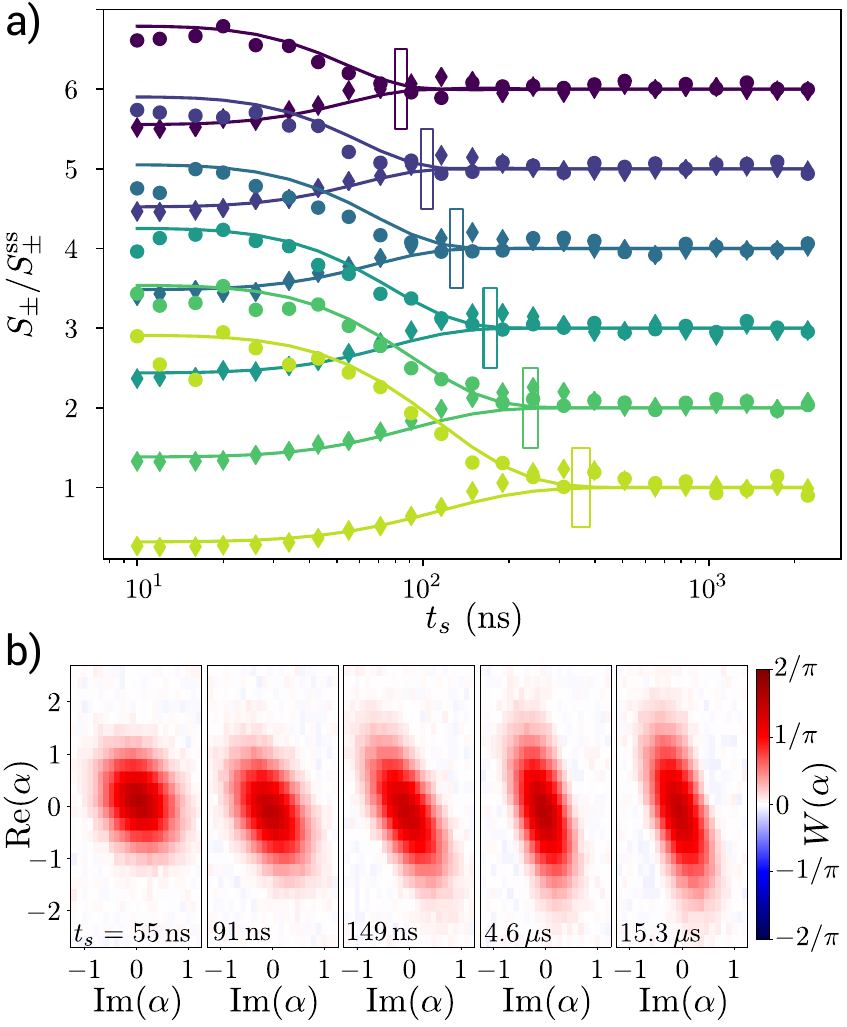}
\caption{Dynamics of the squeezing factors. a) Measured squeezing (dots) and anti-squeezing (diamonds) factors normalized by their steady state values $S_\pm^\mathrm{ss}$ as a function of the stabilization time $t_\mathrm{s}$ for $g_+/2\pi = \SI{1.16}{MHz}$ and various $g_-$. The data are shifted by $1$ for each value of $g_-$ ranging in $g_-/2\pi =$ [1.48, 1.85, 2.22, 2.59, 2.96, 3.33]  \SI{}{MHz} (from light green to dark blue).
Solid lines: results of the numerical simulation. Rectangles indicate the predicted characteristic stabilization times $\kappa_d/\mathcal{G}^2$ assuming \SI{2}{\%} relative uncertainty on $g_+$ and $g_-$. 
b) Selected measured Wigner tomograms for $g_-/2\pi = \SI{1.85}{MHz}$ and $g_+/2\pi = \SI{1.16}{MHz}$ after various stabilization times $t_\mathrm{s}$.} 
\label{fig:sqz_dynamic}
\end{figure}

Our measurements establish that, as expected, the reservoir engineering scheme we implement is able to stabilize a squeezed state in the cavity. In addition to characterizing the steady state, it is also interesting to ask how long the scheme takes to prepare the steady state.
For the ideal (Kerr-free) system, and in the limit of a large dump-mode damping, one can use adiabatic elimination to show that this preparation timescale is   
$\kappa_d/\mathcal{G}^2$~\cite{Kronwald2013}. 

We can directly test this prediction in our experiment.
The measured squeezing and anti-squeezing factors are shown in \cref{fig:sqz_dynamic}.a) as a function of the time $t_\mathrm{s}$ during which the pumps are turned on for $g_+/2\pi = \SI{1.16}{MHz}$ and for various values of $g_-$. By normalizing the squeezing and anti-squeezing factors $S_\pm$ by their steady-state values $S_\pm^\mathrm{ss}$, we observe, as expected, that the steady-state is reached in a typical time of $\kappa_\mathrm{d}/\mathcal{G}^2$ that decreases with $g_-$ (rectangles in \cref{fig:sqz_dynamic}.a). As a consequence, the stabilization time increases with squeezing when considering a fixed $g_+$ value, as long as $\mathcal{G}$ dominates both the cavity loss rate $\kappa_c$ and its self-Kerr rate $K$. This is well-understood from the cooling dynamics of the Bogoliubov mode: larger squeezing parameters $r = \tanh^{-1}(g_{\rm +}/g_{\rm -})$ are obtained for smaller values of $\mathcal{G} = \sqrt{g_{\rm -}^2 - g_{\rm +}^2}$ but they lead to a longer relaxation time. The evolution of the squeezing factors is reproduced using numerical simulation of the master equation (solid lines in \cref{fig:sqz_dynamic}.a). 

It is also interesting to examine experimentally the time-evolution of the full cavity Wigner functions. In \cref{fig:sqz_dynamic}.b), the evolution at $g_+/2\pi = \SI{1.16}{MHz}$ and $g_-/2\pi = \SI{1.85}{MHz}$ (global minimum of the squeezing factor) shows how the squeezing establishes with some rotation and distortion of the Gaussian distribution due to Kerr effect as the average number of photons gets larger. 

\begin{figure}
\centering
\includegraphics[width=8.6cm]{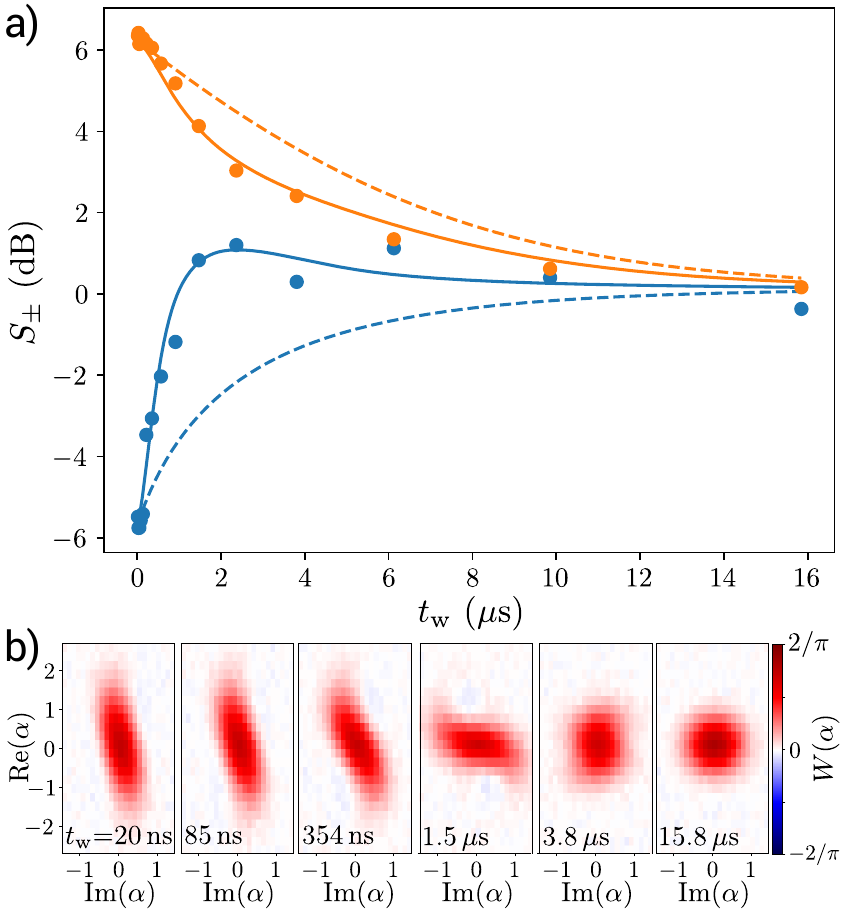}
\caption{Decay of the squeezed state towards thermal equilibrium. a) Dots: measured squeezing factor $S_-$ (blue) and anti-squeezing factor $S_+$ (orange) as a function of the waiting time $t_\mathrm{w}$ during which the pumps are turned off after they were at $g_+/2\pi = \SI{1.42}{MHz}$ and $g_-/2\pi = \SI{2.6}{MHz}$. Solid lines: numerical simulations using $K/\kappa_\mathrm{c} = 0.5$. Dashed lines: same simulations but without self-Kerr ($K=0$). b) Measured Wigner functions after various pump off times $t_\mathrm{w}$ from \SI{20}{ns} to \SI{15.8}{\mu s} as indicated on each label. } 
\label{fig:sqz_relax}
\end{figure}

The steady-state is thus reached in about $\kappa_\mathrm{d}/\mathcal{G}^2$ but how fast does it disappear once the pumps are turned off?
Operating at $g_+/2\pi = \SI{1.42}{MHz}$ and $g_-/2\pi = \SI{2.6}{MHz}$, we perform a Wigner tomography and compute the squeezing and anti-squeezing after a waiting time $t_\mathrm{w}$ (\cref{fig:sqz_relax}.a). A fast decrease of the squeezing factor is observed in a characteristic time shorter than the cavity relaxation time $\kappa_\mathrm{c}^{-1}$. We attribute this deviation from the behavior expected of a perfectly harmonic oscillator (dashed lines) to the self-Kerr effect induced by the transmon qubit onto the cavity. The corresponding predicted evolution of squeezing factors is shown with $K/\kappa_\mathrm{c} = 0.5$ as solid lines in \cref{fig:sqz_relax}.a). In numerical simulations, we observe a transition from over-damped to under-damped oscillations of the squeezing factor $S_-$ as $K/\kappa_c$ increases beyond about $1$ (\cref{append:sqz_relax_vs_K}). Since $K \simeq \kappa_\mathrm{c}$ in the experiment, we are close to a critical damping regime.

\section{Conclusion}

Using dissipation engineering, we have shown the stabilization of a squeezed state in a microwave resonator with a squeezing factor greatly exceeding the standard 3~dB limit for  coherent \textit{in-situ} parametric pumping. We directly measure the squeezing factor by performing a direct Wigner tomography using an ancillary qubit. Correcting for state evolution during measurement, we infer that we achieve a squeezing factor of \SI{-8.2 \pm 0.8}{dB}. While reservoir-engineered squeezing of mechanical modes has previously been demonstrated, this is the first demonstration of this method (to our knowledge) in an electromagnetic system.  The 
reservoir engineering technique used here thus extends the state-of-the-art for intra-resonator microwave squeezing. Moreover, the produced squeezed state is close to a pure state with purity of \SI{-0.4 \pm 0.4}{dB}. 
A displaced vacuum squeezed state could also be stabilized in our system by adding a coherent drive on the dump. 

Beyond the stabilization of Gaussian squeezed states, the techniques presented here could be useful for the stabilization of far more complex states.  As discussed, Kerr nonlinearities already play an appreciable role in our experiment.  Future work could use this nonlinearity directly as a resource for non-Gaussian state preparation.  Recent work has demonstrated that the combination of squeezing-via-parametric driving with Kerr interactions can be used to generate cat states \cite{Goto2016,Puri2017} and even entangled cat states \cite{Mamaev2018}.  The combination of dissipative squeezing (as realized here) with Kerr interactions could similarly yield complex cat-like states.  Our techniques could also be used to generate squeezed Fock states~\cite{Kienzler2017}, squeezed Schrödinger's cat states~\cite{Hsiang-Yu2015} or for the preparation of grid states without the need for measurement~\cite{Fluhmann2019,Hastrup2020,Neeve2020,Campagne-Ibarcq2020}.
These engineered squeezed states could find many applications. Indeed, used to erase which-path information, they can increase gate fidelity~\cite{Puri2016}; used to increase distinguishability, they can improve qubit state readout~\cite{Didier2014, Eddins2018, Didier2015, Touzard2019}. Squeezing can also be used in spin detection to enhance the light-matter coupling~\cite{Leroux2018, Qin2018}. Finally, dissipative squeezing techniques employed on a single site of a lattice of microwave resonators (see e.g.~Ref.~\cite{Owens2018}) can serve as a shortcut for effectively generating highly-entangled many-body states \cite{Yanay2018,Yanay2020}.  

\begin{acknowledgments}
We are grateful to Olivier Arcizet and Alexandre Blais for discussions. This work was initiated during a discussion that happened during Les Houches Summer School in July 2019.
We acknowledge IARPA and Lincoln Labs for providing a Josephson Traveling-Wave Parametric Amplifier. The device was fabricated in the cleanrooms of Collège de France, ENS Paris, CEA Saclay, and Observatoire de Paris. This work is part of a project that has received funding from the European Union's Horizon 2020 research and innovation program under grant agreement No 820505.  AC acknowledges support from the Air Force Office of Scientific Research MURI program, under Grant No. FA9550-19-1-0399. 
\end{acknowledgments}

\appendix
\section{Steady-state Wigners tomograms}
The Wigners tomograms of all the points of \cref{fig:steadystate} are available on~\cite{Wigners_on_zenodo}. 

\section{Sample and setup}
\label{append:setup}
\begin{figure*}[h]
    \centering
    \includegraphics[width=16cm]{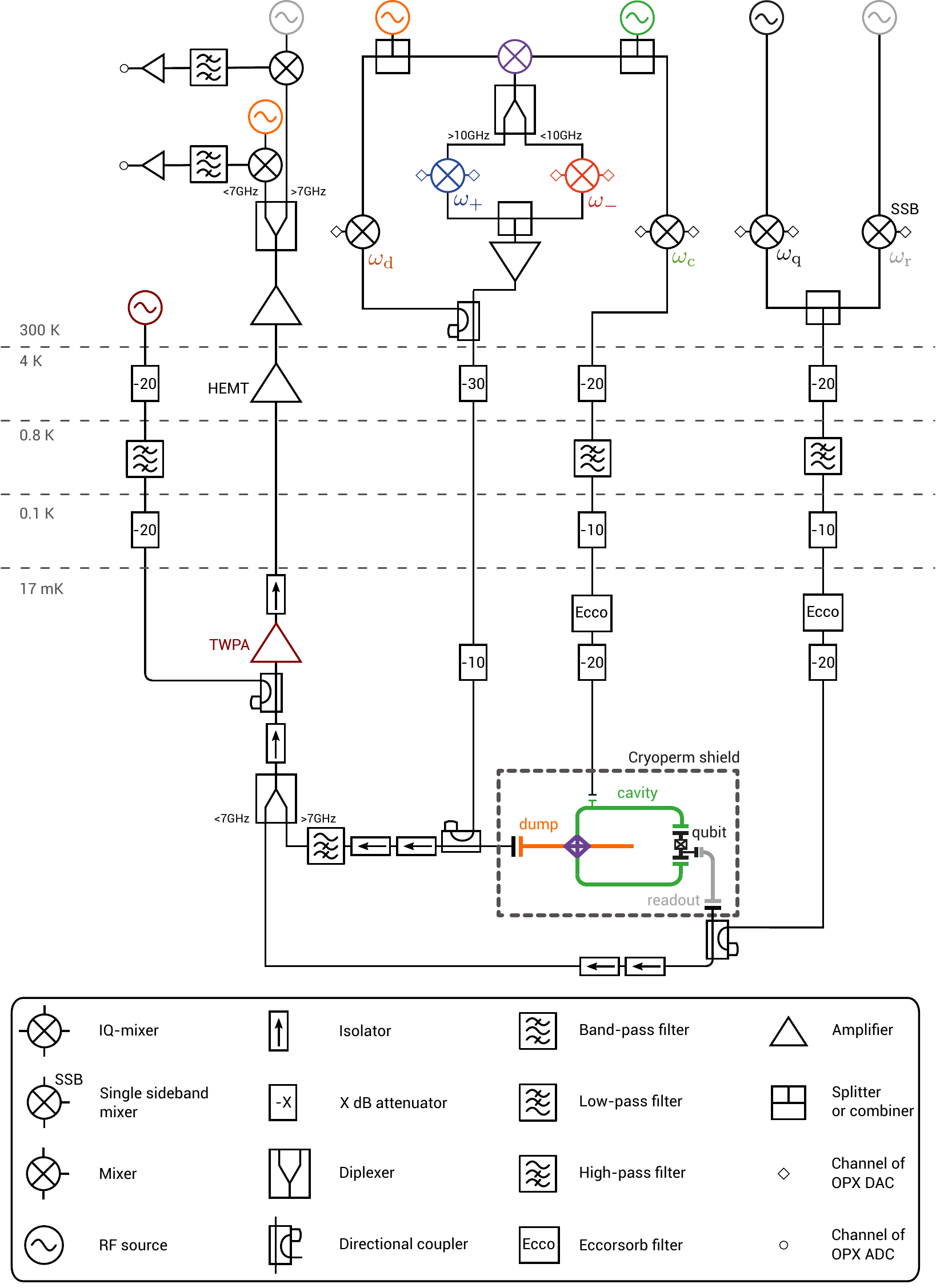}
    \caption{Schematic of the measurement setup. The rf sources color refers to the frequency of the  matching element in the device up to a modulation frequency. Multiple instances of a microwave source with the same color represent a single instrument with split outputs. The sum (blue) and difference (red) pumps are obtained by mixing the cavity and dump rf sources to ensure phase stability. The TWPA~\cite{Macklin2015} was provided by Lincoln Labs.
    \label{fig:wiring}}
\end{figure*}

The sample is the same as in Ref.\cite{dassonneville2020b} albeit for a different cool-down. The measurement setup is also similar with the addition of the pump at the sum frequency (\cref{fig:wiring}). 
The two local oscillators for the pumps are generated by mixing the output of the two microwave sources that are used to generate the dump and cavity drives. 
Intermediate frequency (IF) signals -- tens of MHz -- generated by the Quantum Machines' OPX hardware are upconverted by these local oscillators. Finally, we combine and amplify the two pumps before combining them to the dump port inside of the dilution refrigerator.

To successfully stabilize and measure a squeezed state on a well-defined squeezing axis ($g_\pm$ real), a good phase coherence is required between the pumps and cavity drives. Our setup ensures this condition by deriving the pumps from the dump and cavity local oscillators. One difficulty of our experiment is the large power required for the pumps to reach maximal squeezing factor. This requires the use of a room-temperature amplifier after the mixers (\cref{fig:wiring}). This amplifier has a slow temperature-induced drift in gain, leading to a relative error of \SI{2}{\%} on the pump amplitudes (corresponding to the horizontal errorbars in \cref{fig:steadystate}.b).

\subsection{Calibration of the pumps}
\label{append:pumps_calib}

This section shows how to relate the IF amplitudes $\mathcal{A}_{-}$ and $\mathcal{A}_{+}$ to the rates $g_{-}$ and $g_{+}$.

\begin{figure}
\centering
\includegraphics[width=8.6cm]{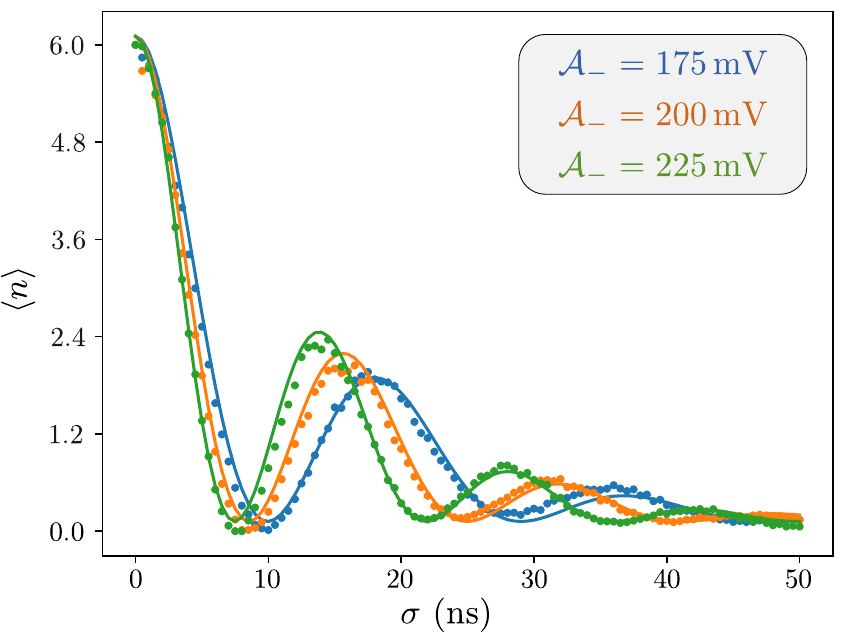}
\caption{Calibration of the rate $g_-$. Dot: measured mean photon number in the cavity as a function of pump pulse width $\sigma$ with a hyperbolic secant shape for three amplitudes $\mathcal{A}_-$. Solid lines: prediction of the photon number using a master equation using $g_-/ \mathcal{A}_-=\SI{74}{MHz \per V}$.}
\label{fig:calib_swap}
\end{figure}

To calibrate $g_-$, we measure the mean photon number in the cavity after applying the pump when the cavity is initially populated with a coherent state $\alpha=\sqrt{6}$. Depending on the amplitude $\mathcal{A}_{-}$ and duration $4\sigma$ of the pump pulse, the rate at which the cavity coherently exchanges excitations with the dump varies (\cref{fig:calib_swap}). Due to the large dissipation rate of the dump, the oscillations of the cavity mean photon number $\expval{n}$ are damped. By fitting the oscillations using a master equation, we find, as expected, a linear dependence of $g_-$ as a function of $\mathcal{A}_{-}$ that we use as calibration.   

\begin{figure}
\centering
\includegraphics[width=8.6cm]{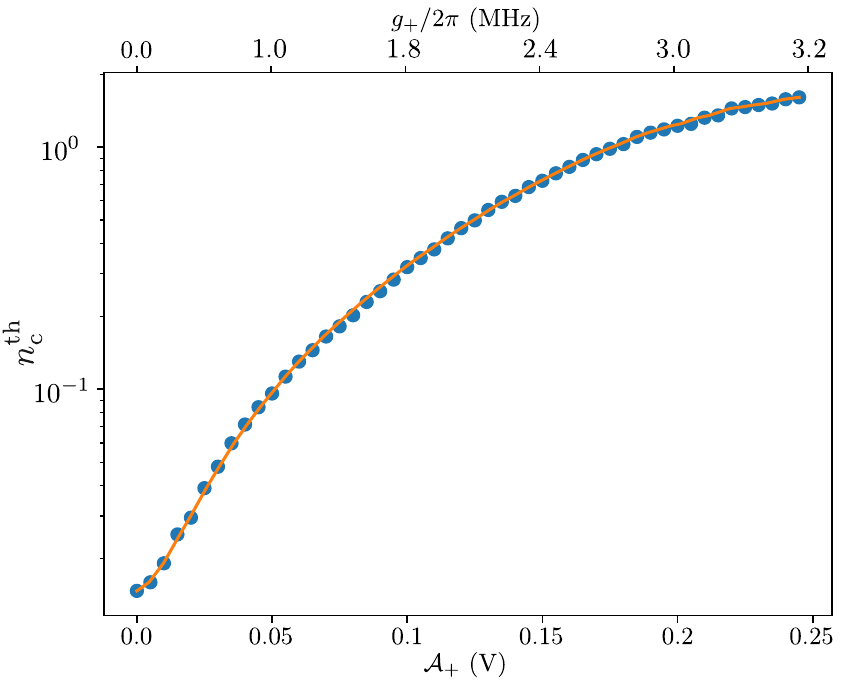}
\caption{Calibration of the rate $g_+$. Dots: measured mean photon number $n^\mathrm{th}_\mathrm{c}$ as function of the amplitude $\mathcal{A}_+$. Numerical simulations allow us to extract the rate $g_{+}$ (top axis) that leads to a given $n^\mathrm{th}_\mathrm{c}$ (see text). Solid line: third order polynomial fit of $g_+$ as function of $\mathcal{A}_+$ that is used as an empirical calibration.}
\label{fig:calib_ampli}
\end{figure}

To calibrate $g_+$, we measure the mean photon number in the cavity $n^\mathrm{th}_\mathrm{c}$ after applying a square pulse with amplitude $\mathcal{A}_+$ for \SI{100}{ns} when the cavity is initially in vacuum. The mean photon number is measured via cavity-induced Ramsey oscillations~\cite{dassonneville2020b}. The only difference with the former reference is that the distribution of photon numbers is thermal instead of Poissonian. Hence, the phase acquired by the qubit during the waiting time of the Ramsey sequence differs and leads to a final qubit excitation probability of
\[
    P_e(t) = \frac{n^\mathrm{th}_\mathrm{c} (1-\cos \chi t) + 1}{2 (1 - \cos \chi t) (n^\mathrm{th}_\mathrm{c} + 1)n^\mathrm{th}_\mathrm{c} + 1} e^{-\Gamma_2 t}.
\]
Using a time dependent master equation, the mean photon number is converted into a two-mode squeezing rate $g_+$. 
The curve $g_+$ as function of $\mathcal{A}_+$ is non linear (\cref{fig:calib_ampli}), likely due to higher order non-linearities in the Hamiltonian. The calibration $g_+(\mathcal{A}_+)$ is then obtained by interpolating the measurement.

\subsection{Cavity displacement calibration}
The calibration of the displacement of the cavity under a pulsed coherent drive is performed by counting the mean photon number. The method chosen to count the mean photon number is to use the ancillary qubit and readout as a vacuum detector~\cite{dassonneville2020b}. This method also allows us to extract the cavity decay rate.

\subsection{Cavity thermal population}
\label{append:calib_nth}
The cavity thermal population is extracted from the cavity-induced Ramsey oscillations of the ancillary qubit~\cite{dassonneville2020b}. 
With the reset protocol, consisting of a swap pulse ($g_-$) between cavity and dump modes followed by a measurement-based feedback initialisation of the qubit in its ground state, we measured a mean photon number $n^\mathrm{th}_\mathrm{c}=1.7 \pm 0.3 \cdot 10^{-2}$ corresponding to an effective temperature of \SI{44 \pm 2}{mK} for the cavity.

\subsection{Correction and uncertainty on the quadrature variances}

We wish to extract the squeezing and anti-squeezing factors by normalizing the measured variances to the zero-point fluctuations. However, the residual thermal population offsets the measured value of the zero-point-fluctuations by a factor $2n_{th} + 1$. This also means that all the measured squeezing factors have to be offset by \SI{0.15 \pm 0.03}{dB}. Due to other sources of uncertainty, such as fluctuations on the cavity displacement pulses, 
we measure a higher statistical uncertainty for the pump-off variances of \SI{\pm 0.2}{dB}.

\section{Kerr-free analytical model}
\label{append:theory_sqz}
This derivation, which can be found in Ref. \cite{Lei2016}, is given here for completeness. When continuously pumping at the difference and sum of the resonance frequencies with rates $g_-$ and $g_+$, the Langevin equations read
\begin{align}
\dot{\hat{d}} = -\frac{\kappa_\mathrm{d}}{2} \hat{d} + i(g_- \hat{c} + g_+ \hat{c}^\dag) + \sqrt{\kappa_\mathrm{d}} \hat{d}_\mathrm{in} \notag \\
\dot{\hat{c}} = -\frac{\kappa_\mathrm{c}}{2} \hat{c} + i(g_- \hat{d} + g_+ \hat{d}^\dag) + \sqrt{\kappa_\mathrm{c}} \hat{c}_\mathrm{in},
\end{align}
where the cavity (dump) input field operators $\hat{c}_\mathrm{in}$ ($\hat{d}_\mathrm{in}$) verify $[\hat{b}_\mathrm{in}(t), \hat{b}_\mathrm{in}(t') ] = \delta (t-t') $ and $\expval{\hat{b}^\dag_\mathrm{in}(t) \hat{b}_\mathrm{in}(t')} = n_\mathrm{b}^\mathrm{th} \delta (t-t')$ for $b=c, d$.
Solving the Langevin equations for the steady-state, the squeezing $S_-$ and anti-squeezing $S_+$ factors are given by
\begin{align}
\label{eq:Kerr-free_analytic}
    S_\pm =& \frac{4(g_- \mp g_+)^2 \kappa_\mathrm{d} (2n_\mathrm{d}^\mathrm{th}+1)}{(\kappa_\mathrm{d}+\kappa_\mathrm{c})(4\mathcal{G}^2 + \kappa_\mathrm{d}\kappa_\mathrm{c})} \notag \\
    +& \frac{ [4\mathcal{G}^2 + \kappa_\mathrm{d}(\kappa_\mathrm{d}+\kappa_\mathrm{c})]\kappa_\mathrm{c}(2n_\mathrm{c}^\mathrm{th}+1) }{(\kappa_\mathrm{d}+\kappa_\mathrm{c})(4\mathcal{G}^2 + \kappa_\mathrm{d}\kappa_\mathrm{c})}.
\end{align}
Assuming $\mathcal{G}, \kappa_\mathrm{c} \ll \kappa_\mathrm{d}$, \cref{eq:Kerr-free_analytic} gives the simplified \cref{eq:analytical_variance} given in the main text.

\section{Modeling the Kerr effect}

The Kerr effect is not included in the analytical model described in \cref{append:theory_sqz}. It induces spurious effects, which reduce the maximal squeezing factor and accelerate the relaxation of squeezing. In this section, we show how to take these effects into account. We simulate our system using the QuantumOptics.jl library~\cite{Julia}. The steady-state simulations are run on an Nvidia Geforce 1080Ti GPU, which allows us to reach Hilbert space dimensions of about 1800. All of the other simulations are run on the CPU. Except for the Wigner tomography retro-prediction, the qubit is not simulated but we take into account the Kerr effect it induces on the cavity. In the case of the Wigner tomography retro-prediction, the dump is adiabatically eliminated.

\subsection{Retro-prediction of the Wigner tomography}
\label{append:wigner_evolution}
\begin{figure*}
\centering
\includegraphics[width=17.2cm]{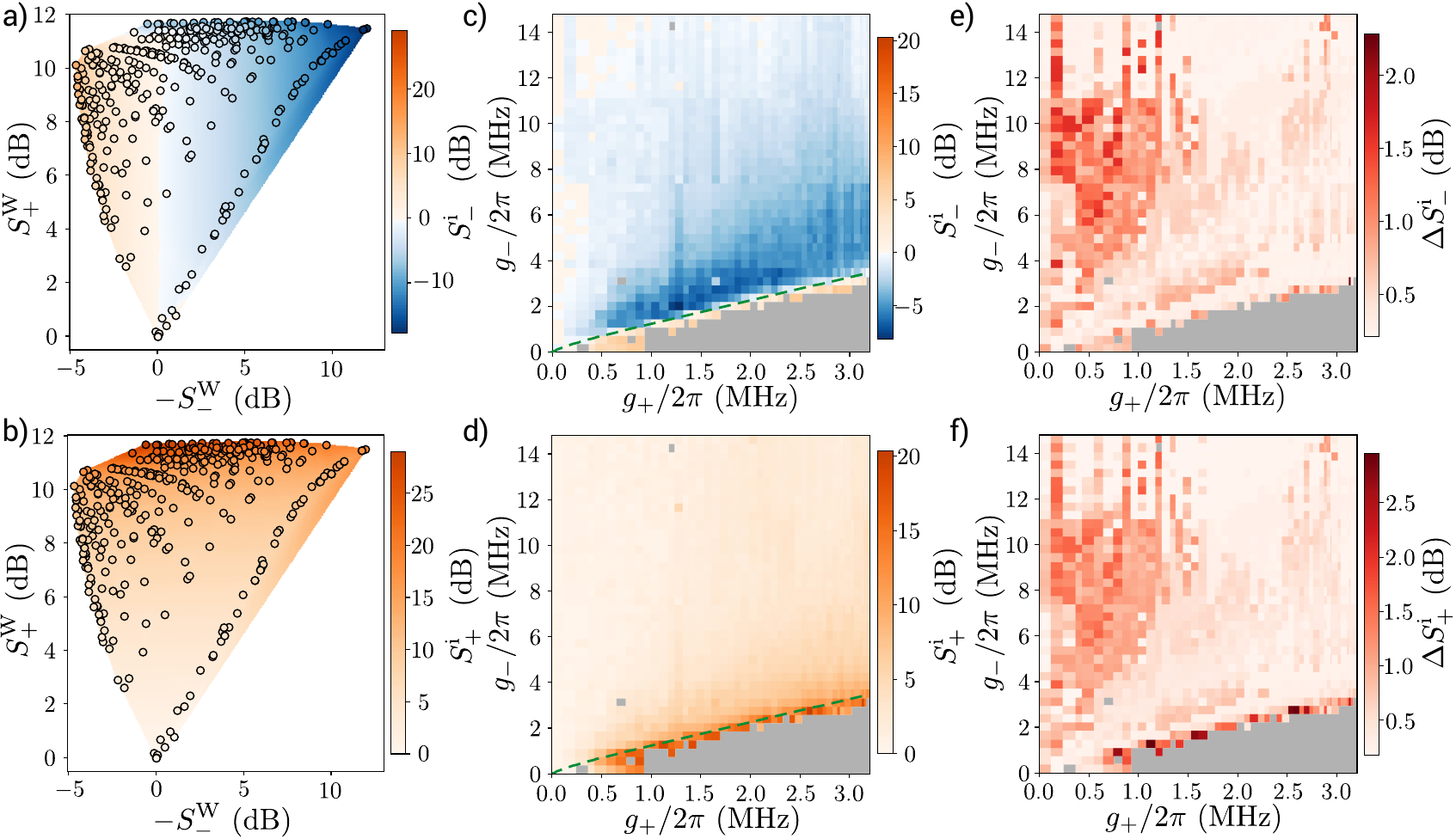}
\caption{a-b) Representation of the map $f_W^{-1}$. Pre-measurement squeezing $S_-^\mathrm{i}$ and anti-squeezing $S_+^\mathrm{i}$ as a function of the measured squeezing and anti-squeezing factors $S_\pm^\mathrm{W}$. Circles: simulated values. Colors: linear interpolation of $f_W^{-1}$.  c-d) Version of \cref{fig:steadystate}.a corrected for the measurement error during Wigner tomography. e-f) Color: retro-predicted uncertainty on the squeezing factors $\Delta S_-^\mathrm{i}$ and $\Delta S_+^\mathrm{i}$ owing to a measurement uncertainty of \SI{\pm 0.2}{dB} on $S_\pm^\mathrm{W}$.}
\label{fig:interpolated_retroprediction}
\end{figure*}
In order to correct for the error introduced by the cavity evolution during Wigner tomography, we resort to simulations of the cavity and qubit alone. Indeed, in the absence of pumps, the effect of the JRM on the cavity is negligible. We numerically implement our experimental Wigner tomography pulse sequence on a truncated Hilbert with up to 50 excitations for the cavity and the two qubit states. Starting from a range of initial squeezed states for the cavity, with variances ($S_-^\mathrm{i}$, $S_+^\mathrm{i}$), we simulate the outcome of the faulty Wigner tomography by computing the variances ($S_-^\mathrm{W}$, $S_+^\mathrm{W}$) of the simulated Wigner tomograms. 

This data-set provides a function $f_W$ that maps actual variances ($S_-^\mathrm{i}$, $S_+^\mathrm{i}$) of the pre-measured quantum state to the variances ($S_-^\mathrm{W}$, $S_+^\mathrm{W}$) extracted from the measured Wigner tomograms. 
As this function empirically appears bijective, the retro-prediction is performed by interpolating its inverse $f_W^{-1}$. The interpolated $f_W^{-1}$ for the initial squeezing and anti-squeezing, as well as the simulated points, are shown in \cref{fig:interpolated_retroprediction}.a and b respectively. 
The retro-predicted initial squeezing and anti-squeezing corresponding to \cref{fig:steadystate}.a-b are shown in \cref{fig:interpolated_retroprediction}.c-d respectively.

Assuming the \SI{\pm 0.2}{dB} of uncertainty on the measured $S_\pm^\mathrm{W}$, and retro-predicting the evolution during Wigner tomography, we obtain an uncertainty $\Delta S_\pm$ on the retro-predicted squeezing and anti-squeezing factors that depends on the value of the measured squeezing and anti-squeezing (\cref{fig:interpolated_retroprediction}.e-f).

\subsection{Steady-state simulations}
\label{append:num_model}
\begin{figure*}
    \centering
    \includegraphics[width=17.2cm]{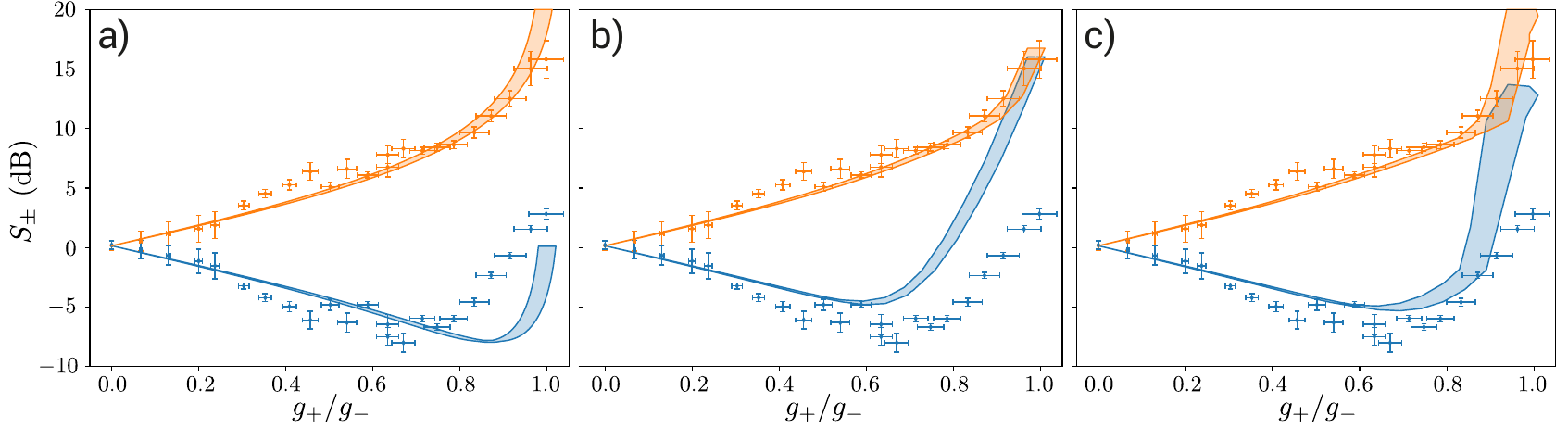}
    \caption{a, b and c) Crosses, retro-predicted squeezing factors for $g_-/2\pi= \SI{1.85}{MHz}$ as in \cref{fig:steadystate}.b. Shaded areas correspond to different models with \SI{2}{\%} uncertainty on $g_+$ and $g_-$;
    in a), Kerr-free analytical model, in b), steady-state simulations with $K/2\pi = \SI{20 \pm 2}{kHz}$, in c), steady state simulations including in addition to the Kerr effect some JRM four-wave mixing terms, $K_\mathrm{cd}/2\pi = \SI{250}{kHz}$, $K_\mathrm{pc} \abs{p_-}^2 /2\pi= \SI{172 \pm 4}{kHz} $ and $K_\mathrm{pd} \abs{p_-}^2 /2\pi= \SI{172 \pm 4}{kHz} $.}
    \label{fig:simu_ss_cut}
\end{figure*}

As seen in \cref{fig:steadystate}.b, the analytical Kerr-free model fails to quantitatively describe the squeezing factor at $g_+ > 0.7 g_-$ (\cref{fig:simu_ss_cut}.a). Here, we compute how higher order terms in the Hamiltonian may  explain this difference.
The first term we consider is the Kerr effect $-K c^{\dag^2} c^2$ induced by the qubit on the cavity (\cref{fig:simu_ss_cut}.b). This simulation accurately predicts the optimal $g_+$ but still fails to reproduce the measured squeezing factors above $g_+/g_- =0.7$. 
Experimentally, we aim for a JRM flux bias that maximizes the three-wave mixing term while cancelling the four-wave mixing term. However, small deviations from this sweet spot create four-wave mixing terms between the cavity, dump and pumps. Contrary to the retroprediction simulations which model a situation where the pumps are turned off, these extra terms may have a significant impact on the squeezing factor where the pumps are turned on.
In the RWA, the four wave-mixing term leads to three kinds of interactions, a cross-Kerr between cavity and dump $K_\mathrm{cd} c^\dag c d^\dag d$, an AC-Stark frequency shift due to the pumps $2(\abs{p_-}^2 + \abs{p_+}^2) ( K_\mathrm{pc} c^\dag c +  K_\mathrm{pd} d^\dag d)$ and parametric squeezing drive due to pump inter-modulation $K_\mathrm{pc} p_+ p_-^* c^{\dag^2}  + K_\mathrm{pd} p_+^* p_-^* d^{\dag^2} + h.c. $. 
The JRM also induces a self-Kerr interaction for the dump, but we neglected it as it is one order of magnitude smaller than $K_\mathrm{pd} p_+^* p_-^*$ in our case \cite{Flurin2014} and much smaller than the dissipation rate $\kappa_d$ anyway. 
The rates $K_\mathrm{cd}$, $K_\mathrm{pc}$ and $K_\mathrm{pd}$ are not measured in this run. Realistic values $K_\mathrm{cd}/2\pi = \SI{250}{kHz}$, $K_\mathrm{pc} \abs{p_-}^2 /2\pi= \SI{172}{kHz} $ and $K_\mathrm{pd} \abs{p_-}^2 /2\pi= \SI{172}{kHz} $ can change the squeezing factors at the large $g_+$, which comforts the assumption that higher order nonlinearities may explain the deviations we observe between our analytical model and the measured squeezing factors.

To numerically compute the steady-state squeezing and anti-squeezing as a function of $g_-$ and $g_+$, we use an iterative method to find the Liouvillian eigenvalues on a truncated Hilbert space comprising up to 60 excitations for the cavity and 30 excitations for the dump.

\subsection{Simulations of the squeezing dynamics}
\label{append:sqz_relax_vs_K}
The dynamics of stabilization and decay of squeezing are computed by solving the master equation on a truncated Hilbert space comprising up to 20 excitations for the cavity and 16 excitations for the dump.

\begin{figure}[h]
    \centering
    \includegraphics[width=8.6cm]{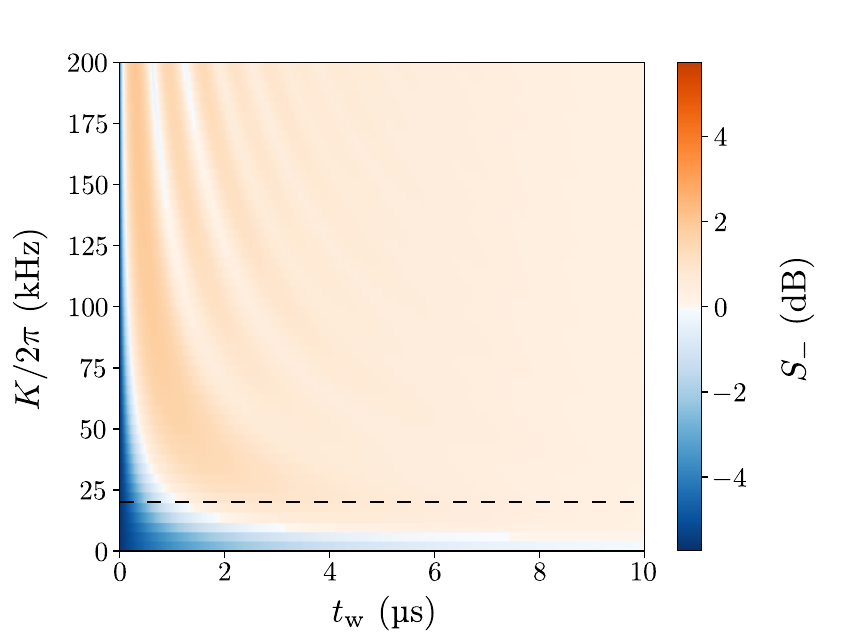}
    \caption{Color: simulated squeezing factor $S_-$ as a function of waiting time $t_\mathrm{w}$ and cavity self-Kerr rate $K$ for an initial state with squeezing factor $S_-=\SI{-5.7}{dB}$ and anti-squeezing of $S_+=\SI{6.2}{dB}$. The dashed line indicates our device parameter $K/2\pi= \SI{20}{kHz}$.}
    \label{fig:simu_Kerr_decay}
\end{figure}

To understand the effect of the self-Kerr term on the squeezing decay, we simulate the evolution of squeezing for varying waiting time $t_\mathrm{w}$ and various self-Kerr rates $K$ (\cref{fig:simu_Kerr_decay}). We initialize the cavity state at $t_\mathrm{w}=0$ in a Gaussian state with the measured $S_-=\SI{-5.7}{dB}$ and $S_+=\SI{6.2}{dB}$ of \cref{fig:sqz_relax}.a. Dismissing the Kerr effect ($K=0$), we observe an exponential damping of squeezing due to the cavity relaxation. For nonzero $K$, the squeezing factor also oscillates in time. Our experimental value $K/2\pi = \SI{20}{kHz}$ is closed to the critically damped regime where the effective decay time is maximally reduced. 
This observation highlights the crucial role of Kerr effect in the imperfections of our Wigner tomography technique used to estimate the variances.

\section{Effect of finite size Wigner tomograms}
\label{append:finite_window}
Due to experimental constraints, the probed quadrature phase space must be finite. A rectangular window  $-x_0\leq\mathrm{Im}(\alpha)\leq x_0$ and $-y_0\leq\mathrm{Re}(\alpha)\leq y_0$ is chosen, where $x_0=1.4$ and $y_0=2.7$.
This induces a systematic error on the estimation of the variances. Indeed, the variances $\expval{X_-^2}$ and $\expval{X_+^2}$ are computed from the Wigner function using $\expval{X_-^2} = \mathrm{min}_\theta \int_{-\infty}^{\infty} \Im(\alpha e^{i\theta})^2 W(\alpha) d^2\alpha $ and $\expval{X_+^2} = \max_\theta \int_{-\infty}^{\infty} \Re(\alpha e^{i\theta})^2 W(\alpha) d^2\alpha $. 
Assuming a vacuum squeezed state with minimal variance $\expval{X_-^2} = \sigma^2$ and maximal variance $\expval{X_+^2} = 1/\sigma^2$, its Wigner function is given by: 
\[ 
    W(\alpha) = \frac{2}{\pi} \exp\left(- \Re(\alpha e^{i\theta_\mathrm{min}})^2 \sigma^2 -\Im(\alpha e^{i\theta_\mathrm{min}})^2 /\sigma^2 \right)
\]
 Knowing only the Wigner function in a window $[-y_0, y_0]$ over the real axis, the inferred maximum variance is 
 \[
    \expval{X_+^2}_{y_0} = \big(-2\sigma y_0 e^{-y_0^2 \sigma^2}/\sqrt{\pi} + \text{Erf}(y_0 \sigma) \big) \frac{1}{\sigma^2} 
\]
This variance saturates as function of $\sigma$ towards a value of \SI{11}{dB} in our case. This means that we are unable to resolve any variance above \SI{11}{dB} along the real axis.


%

\end{document}